# Phase transitions and thermodynamic properties of antiferromagnetic Ising model with next-nearest-neighbor interactions on the Kagomé lattice

M.A. Magomedov [a,b*], M.K. Ramazanov [a,b], A.K. Murtazaev [a]

[a] Institute of Physics, Dagestan Scientific Center of the Russian Academy of Sciences,
367003, Makhachkala, Russia
[b] Dagestan Scientific Center of the Russian Academy of Sciences,
367000, Makhachkala, Russian Federation
[*] e-mail: magomedov_ma@mail.ru

**Abstract.** We study phase transitions and thermodynamic properties in the two-dimensional antiferromagnetic Ising model with next-nearest-neighbor interaction on a Kagomé lattice by Monte Carlo simulations. A histogram data analysis shows that a second order transition occurs in the model. From the analysis of obtained data, we can assume that next-nearest-neighbor ferromagnetic interactions in two-dimensional antiferromagnetic Ising model on a Kagomé lattice excite the occurrence of a second order transition and unusual behavior of thermodynamic properties on the temperature dependence.

## 1. Introduction

There has lately been increased focus on the study of phase transitions (PTs), a critical behavior, and thermodynamic properties in compounds with a Kagomé lattice. Kagomé lattice spin systems are strongly frustrated because of their specific geometry. The ordering in such systems takes place more slowly at decreasing the temperature even in comparison with common frustrated systems. It results from that systems with a lower coordination number can manifest not only the non-trivial global degeneration but also locally degenerated states [1, 2].

In Kagomé spin systems with only nearest neighbor exchange interactions, a magneto-ordered PT is not implemented at any non-zero temperature. Next-nearest-neighbor exchange interactions remove the degeneration and can excite PT at non-zero temperatures [3]. Nevertheless, since frustration effects still take place, the ordering and stabilization of structures are slowed down in contrast to non-frustrated systems [4].

The thermodynamic, magnetic, and critical properties of spin systems as well as the influence of frustrations, appeared at competing interactions caused by the lattice geometry, on PTs are investigated in [5-8]. Authors study the phenomena of the appearance and extinction of frustration in the dependence on an external magnetic field value, signs and value of correlation $r=J_2/J_1$, where $J_1$ and $J_2$ are the constants of nearest and next-nearest-neighbor exchange interactions, correspondingly.

The two-dimensional antiferromagnetic Ising model on a Kagomé lattice, studied intensively over the last years, is the instance of a geometrically frustrated system [9-12]. In the ground state of this model with nearest-neighbor interaction the entropy accounted for per spin is non-zero [13] and hence a PT is absent. This stems from the strong decrease of the spin ordering due to frustration effects. However, next-nearest-neighbor interactions stabilize the spin state and a PT occurs in the system [4, 9]. The next-nearest-neighbor interactions in the classical Ising model are accompanied by the ground state degeneracy, the occurrence of various phases and PTs also can influence on critical and thermodynamic properties, in particular, there are appeared different anomalies of critical properties [14].

Therefore, for better insight into a thermodynamic behavior of systems with competing interactions it is necessary to perform more accurate study of the antiferromagnetic Ising model on a Kagomé lattice with next-nearest-neighbor interactions using additional contemporary ideas and methods.

## 2. Model and method

The Kagomé Ising antiferromagnetic model with next-nearest-neighbor interactions is described by Hamiltonian:

$$H = J_1 \sum_{\langle i,j \rangle} S_i S_j + J_2 \sum_{\langle i,l \rangle} S_i S_l, \quad (1)$$

where $S=\pm 1$ is the Ising spin. A first term in Eq. 1 accounts for the nearest neighbor antiferromagnetic exchange interaction of $J_1>0$, a second term accounts for next-nearest-neighbor ferromagnetic interaction of $J_2<0$. In this study, we consider cases of $r = |J_2/J_1| = 0.5; 1.0; 1.5$, and 2.0.

Such microscopic Hamiltonian-based systems can be strictly and successively investigated using the Monte Carlo simulation [15-19]. Among a good deal of Monte Carlo algorithms developed in recent years, the Wang-Landau algorithm is the most efficient for estimating such systems, especially in a low-temperature region [20, 21].

The Wang-Landau algorithm is the realization of an entropy simulation and used to calculate the density of states of a system. It performs a random walk in energy space with probabilities inverse to the density of states $g(E)$ achieving a uniform energy distribution. The transition probabilities are fitted so that all energy states will be visited, what permits to obtained a first-time-unknown density of states $g(E)$ and compute the values of required thermodynamic parameters observables at any temperature. Since the density of states rapidly rises with an increase in sizes of studied systems, one uses the $\ln g(E)$ value for convenience in storing and processing the large numbers.

In this study, we use the Wang-Landau algorithm as follows. We start with an initial arbitrary configuration of spins: the density of states is $g(E)=1$, the energy distribution histogram is $H(E)=0$, and the initial modification factor is $f = f_0 = e^1 \approx 2.71828$. We proceed with steps in a phase space until a flat histogram $H(E)$ is obtained (i.e. until all possible energy states of a system will be visited an approximately equal number of time). However, the change probability of the state with energy $E_1$ to $E_2$ is determined by formula $p = g(E_1)/g(E_2)$. If the change to $E_2$ occured then $g(E_2) \to f \times g(E_2)$, $H(E_2) \to H(E_2)+1$, otherwise $g(E_1) \to f \times g(E_1)$, $H(E_1) \to H(E_1)+1$.

When the histogram is flat, we reset the histogram to $H(E) \to 0$, reduce the modification factor $f \to \sqrt{f}$, and continue again until $f \geq f_{min}$. In our case $f_{min} = 1.0000000001$. More detailed description of the Wang-Landau algorithm was is reported in [22-25]. Thus, using the data for density of states defined in this fashion we can compute thermodynamic parameter values at any temperature. Particularly, the internal energy $U$, the free energy $F$, and the entropy $S$ can be computed by equations

$$U(T) = \frac{\sum_E E g(E) e^{-E/k_B T}}{\sum_E g(E) e^{-E/k_B T}} \equiv <E>_T, \quad (2)$$

$$F(T) = -k_B T \ln \left( \sum_E g(E) e^{-E/k_B T} \right), \quad (3)$$

$$S(T) = \frac{U(T) - F(T)}{T}. \quad (4)$$

A character of PTs is analyzed using the Monte Carlo histogram technique [26]. We perform computations for systems with periodic boundary conditions and linear sizes $L=12 \div 120$, a number of particles is $N = \frac{3}{4} \times L \times L$.

### 3. Simulation Results

The temperature behavior of the heat capacity and the susceptibility is defined using equations [27, 28]

$$C = (NK^2)(\langle U^2 \rangle - \langle U \rangle^2), \qquad (5)$$

$$\chi = \begin{cases} (NK)(\langle m^2 \rangle - \langle |m| \rangle^2), & T < T_N \\ (NK)\langle m^2 \rangle, & T \geq T_N \end{cases}, \qquad (6)$$

where $K = |J_1|/k_B T$, $N$ is the number of particles, $m$ is the ordering parameter ($U$ and $m$ are normalized quantities).

The order parameter of the system is calculated from equation [9]

$$m = \frac{1}{3}(|m_1| + |m_2| + |m_3|), \qquad (7)$$

where $m_i$ is the magnetization per spin for one of three sublattice.

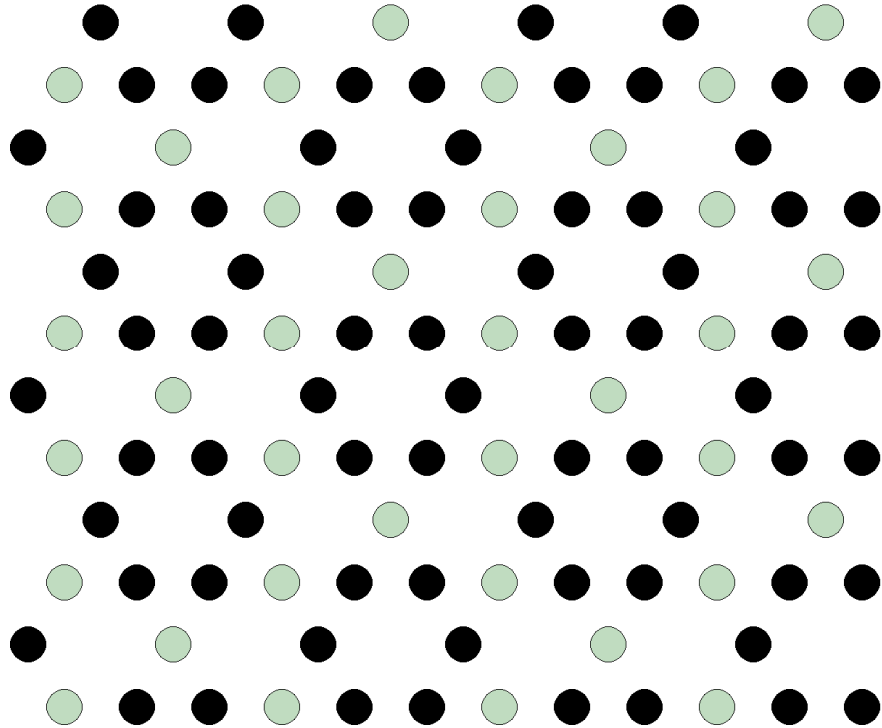

Figure 1. The magnetic structure of the ground state.

Figure 1 presents the magnetic structure of the ground state for the model under study obtained by means of the Wang-Landau algorithm. Black circles denote spins up, and light green circles are for spins down. As is evident from the figure, the ground state possesses a ferrimagnetic ordering.

The density of states for systems with different linear sizes $L$ is shown in Figure 2 (hereafter a statistic error doesn`t exceed symbol sizes used for dependences plotting). As a matter of convenience only a few of data are symbolized, all residue data fall on the line. The diagram depicts that the degeneration of the ground state in this system is absent. We assume that this is due to next-nearest-neighbor exchange interaction which facilitates to partial removal of the degeneration.

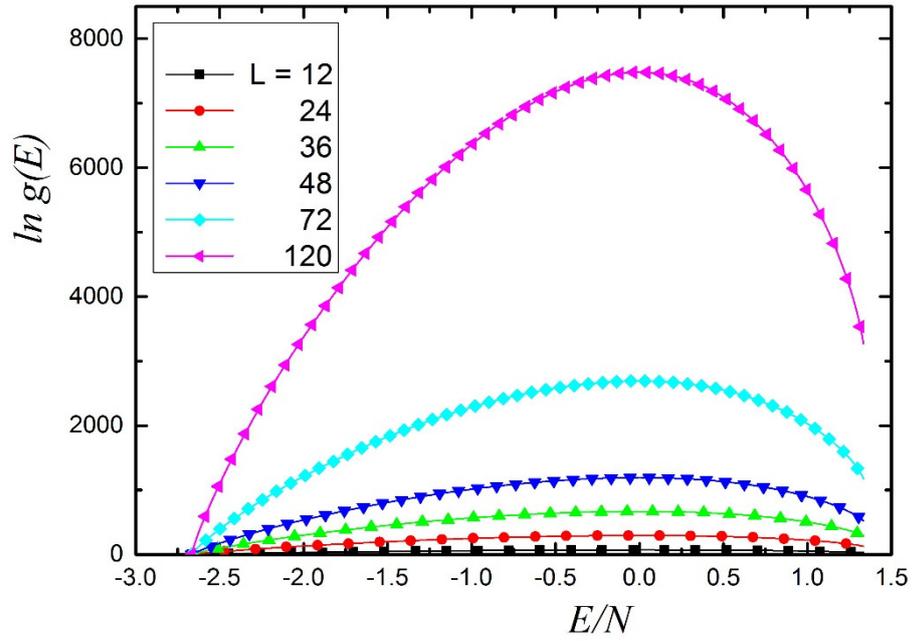

Figure 2. The density of states *g(E)* for systems with different linear sizes *L*.

The entropy temperature dependences *S* at different *L* are depicted in Figs. 3 and 4 (hereafter the temperature is given in $|J_1|/k_B$) units). The system entropy tends to predicted ln2 value at increasing the temperature (Figure 3).

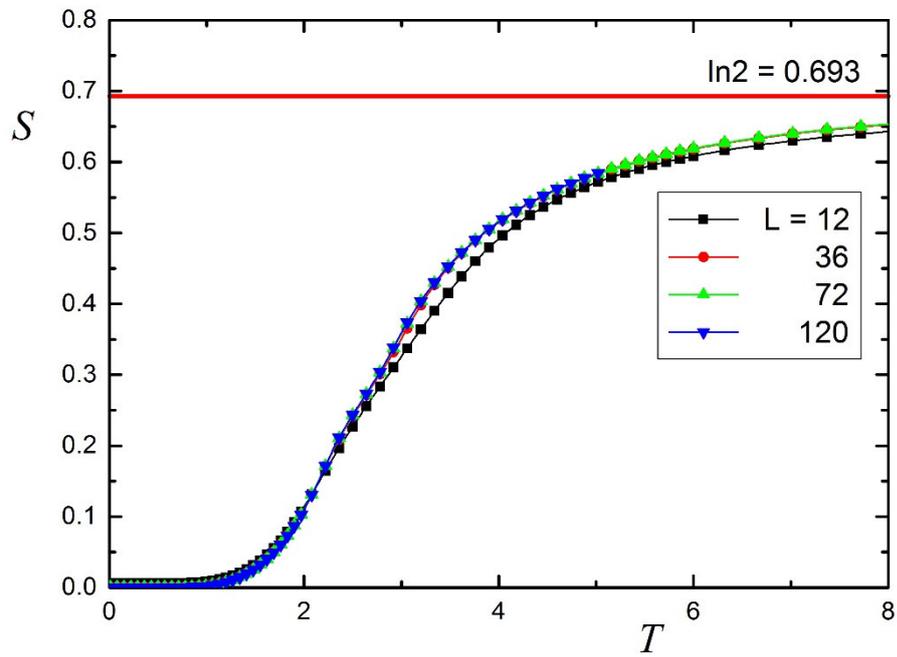

Figure 3. The temperature dependence of the entropy *S*.

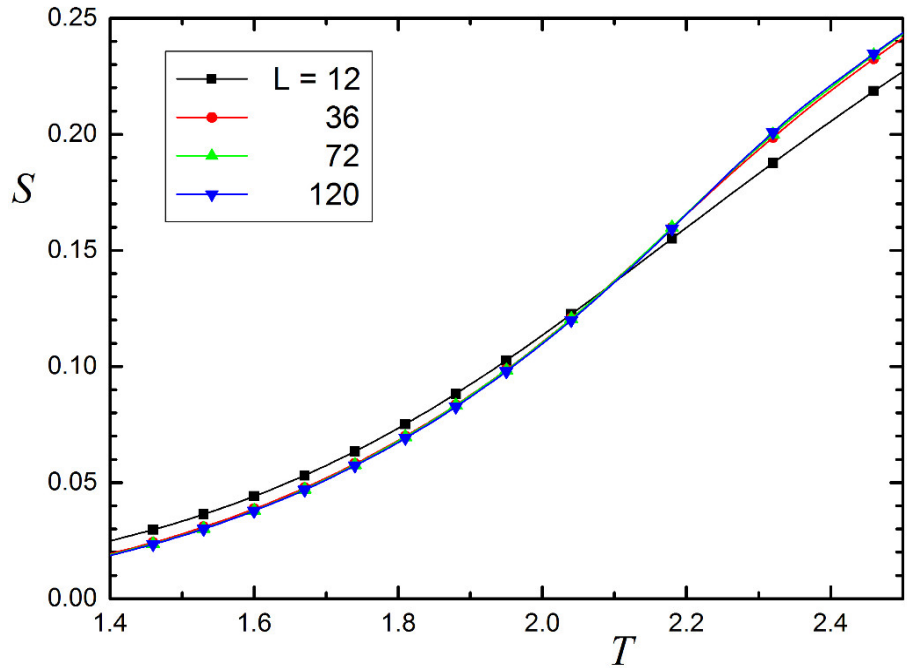

Figure 4. The temperature dependence of the entropy $S$.

At low temperatures close to the absolute zero the entropy goes to zero, while for the model with nearest neighbor interactions that goes to a nonzero value. Moreover, the points of the entropy intersection for systems with different $L$ are found on the entropy temperature dependence, which is distinctive feature of this model (Figure 4).

Figs. 5 and 6 represent the temperature dependences of the heat capacity and susceptibility at different $L$. We note that both the heat capacity and the susceptibility demonstrate an unusual behavior expressed by the presence of double peak. It is evident from diagrams that the absolute values of the heat capacity and susceptibility maxima grow with increase in $L$. In addition, the double-peak structure becomes more pronounced. Such a behavior, obviously, is associated with the competition of nearest and next-nearest-neighbors. It can be assumed that the first maximum on diagrams is caused by a transition from ordered to partially disordered states, and the second one corresponds to the transition of the system into the paramagnetic state.

The temperature dependences of the order parameter $m$ at different $L$ are depicted in Figure 7. As is evident from the figure, the order parameter shows an unusual behavior which is enhanced with the growth in system linear sizes. Such behavior of the system fall on the same values at which two peaks of the heat capacity and susceptibility are found (Figs. 5 and 6).

Figs. 8 and 9 show temperature dependences of the heat capacity and the order parameter at $L = 48$ and $L = 120$ for different $r$ values. Note that the $r$ growth within $0.5 \leq r \leq 2.0$ is accompanied by shifting the maxima towards high temperatures. As can be seen from this Figure 8, a double peak for the heat capacity is observed for the whole interval $0.5 \leq r \leq 2.0$. The same result was obtained for susceptibility. As seen for fig 9 the order parameter decreased to 0.6 on first maximum of heat capacity and vanish on second maximum.

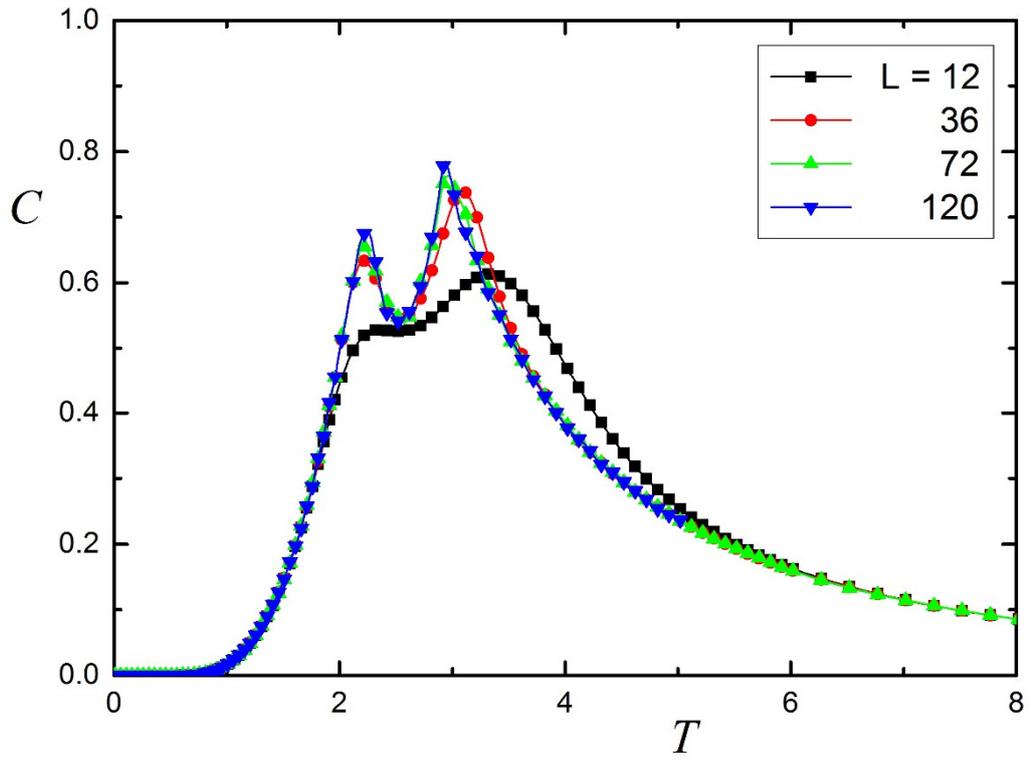

Figure 5. The temperature dependence of the heat capacity *C*.

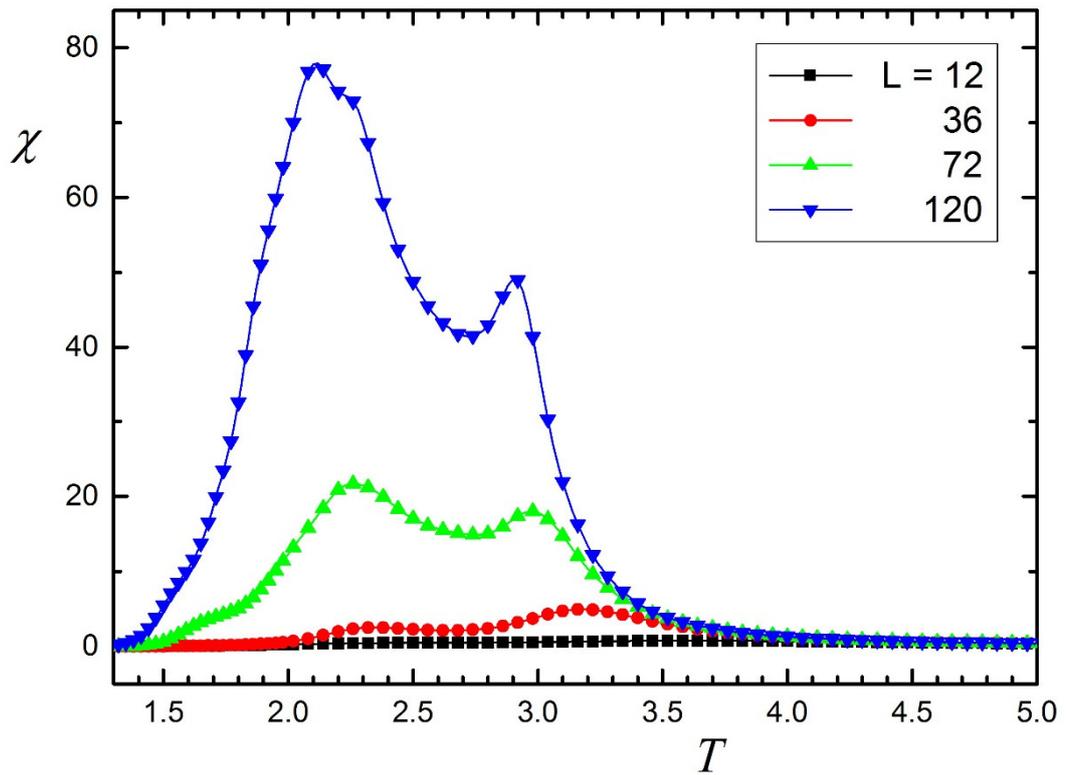

Figure 6. The temperature dependence of the susceptibility $\chi$.

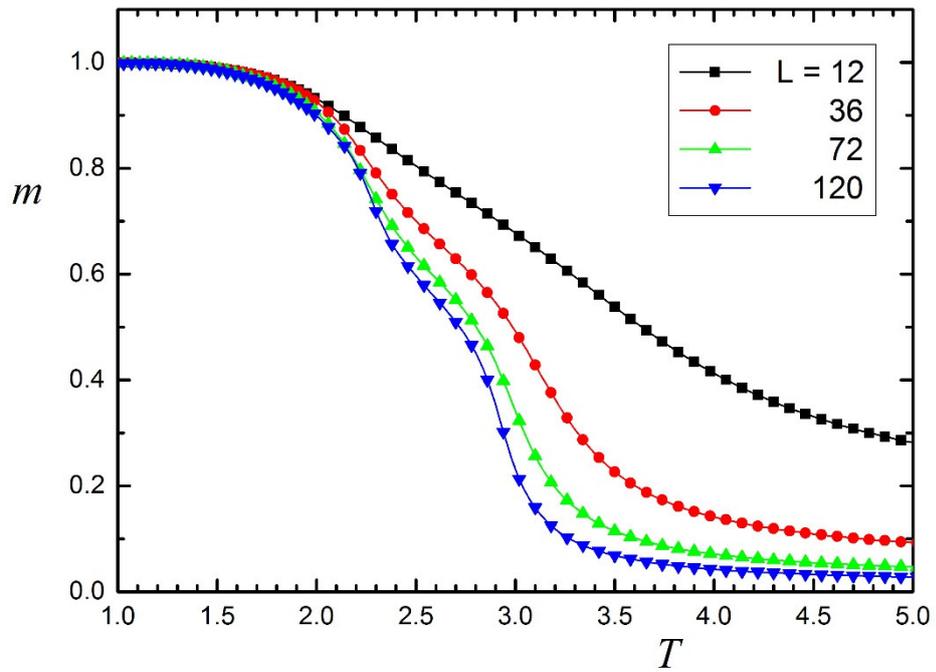

Figure 7. The temperature dependence of the order parameter *m*.

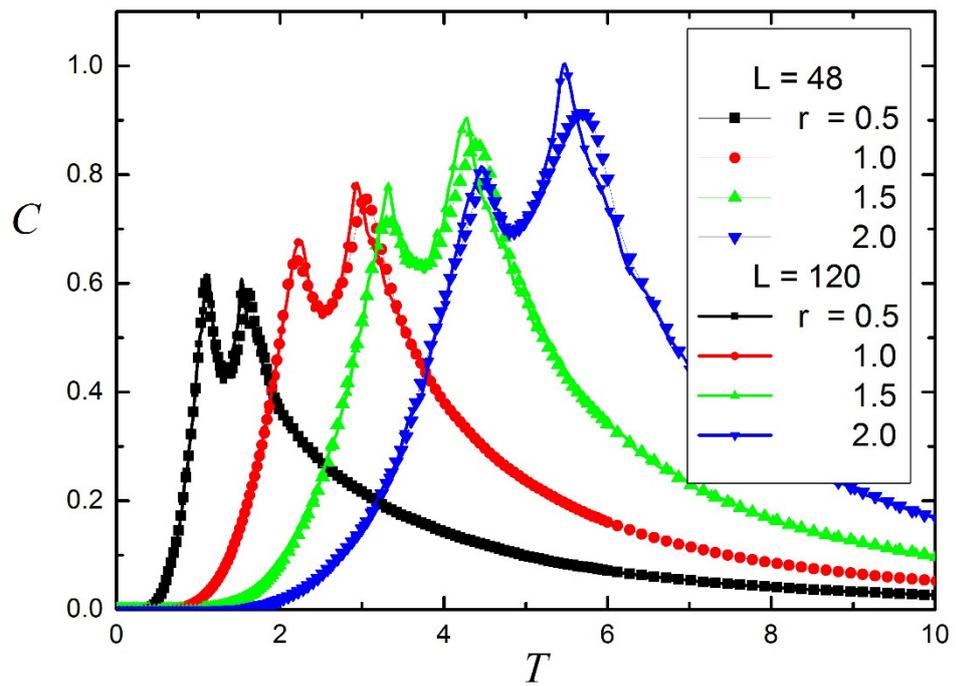

Figure 8. The temperature dependence of the heat capacity *C* for different *r*.

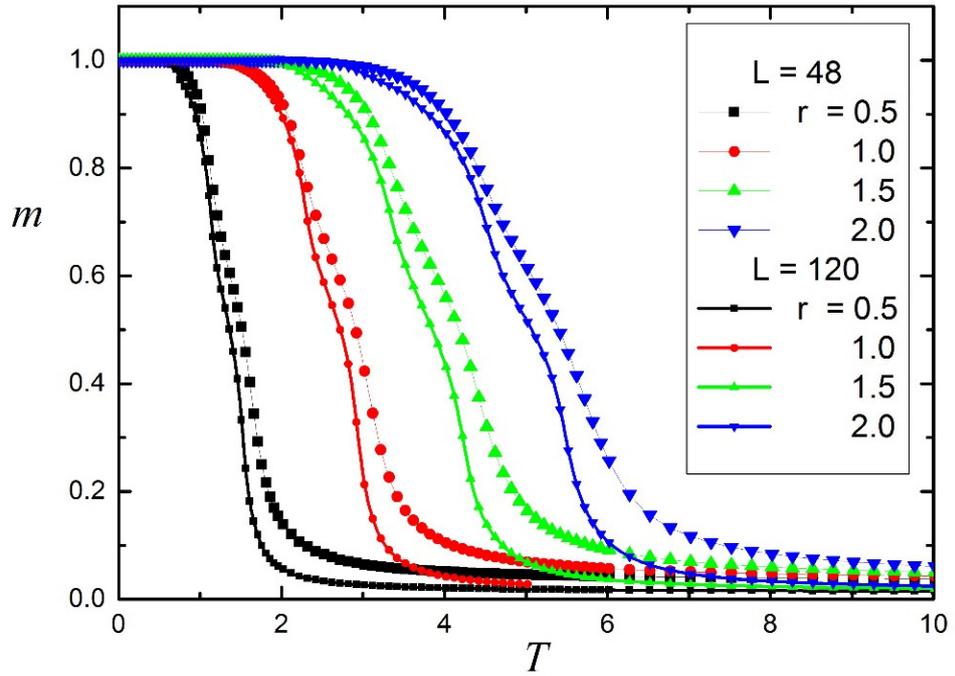

Figure 9. The temperature dependence of the order parameter *m* for different *r*.

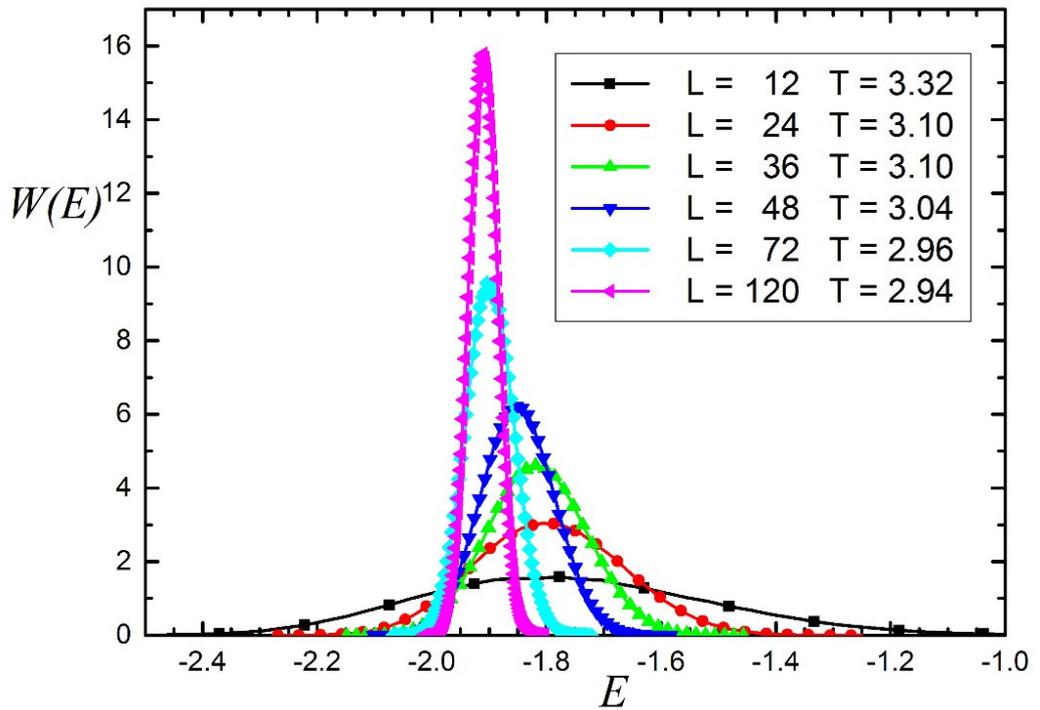

Figure 10. The histogram of the energy distributions *W(E)*.

Figure 10 represents the energy distribution histograms for systems with different *L*. Diagrams are plotted near the point relating to the temperature of the heat capacity second maximum. Because the temperatures of the heat capacity maxima for different *L* vary, corresponding temperature maxima are shown on diagrams. On diagrams we observe one peak that features for a second order transition. From the analysis of obtained data, we can assume that next-nearest-neighbor ferromagnetic interactions in two-dimensional antiferromagnetic Ising model on a Kagomé lattice excite the occurrence of a second order transition and unusual behavior of thermodynamic properties on the temperature dependence.

9## 4. Conclusions

We studied the phase transitions in two-dimensional antiferromagnetic Ising model on a Kagomé lattice with next-nearest-neighbor interactions using the high-performance Monte Carlo Wang-Landau algorithm. The phase transition character was analyzed by the histogram technique. The phase transition occurred in the model under study was determined to be of the second order. We found that the next-nearest-neighbor ferromagnetic interactions in the studied model contribute to the abnormal behavior of thermodynamic parameters found on the temperature dependence.

**Acknowledgments**

The study was supported by grant of RFBR (Project no. 16-02-00214).

## Additional materials

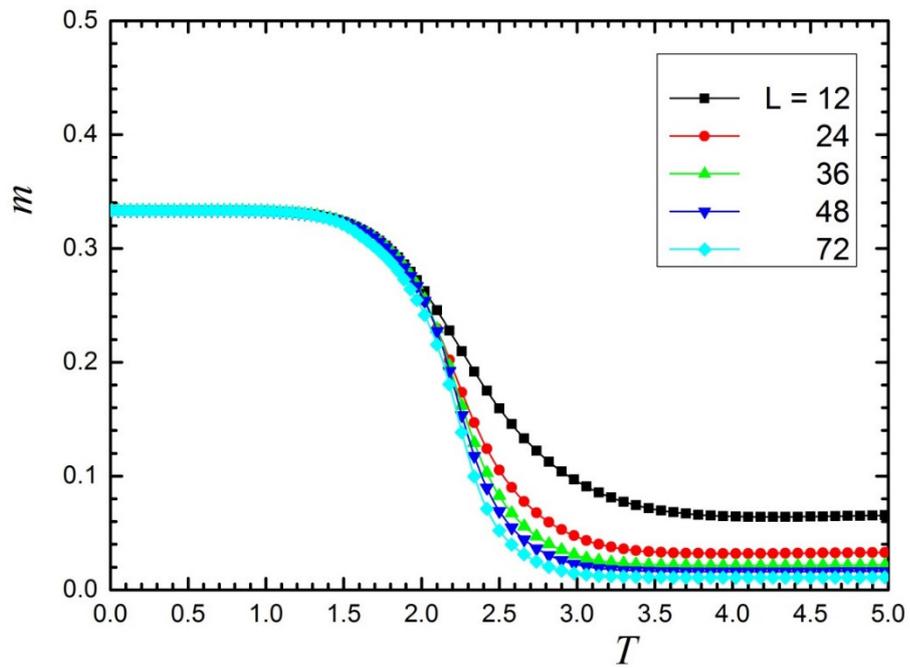

Figure 11. The temperature dependence of the full magnetic moment *m*.

Magnetic moment defined as:

$$\langle m \rangle = \left\langle \left| \sum_i S_i \right| \right\rangle \quad (8)$$

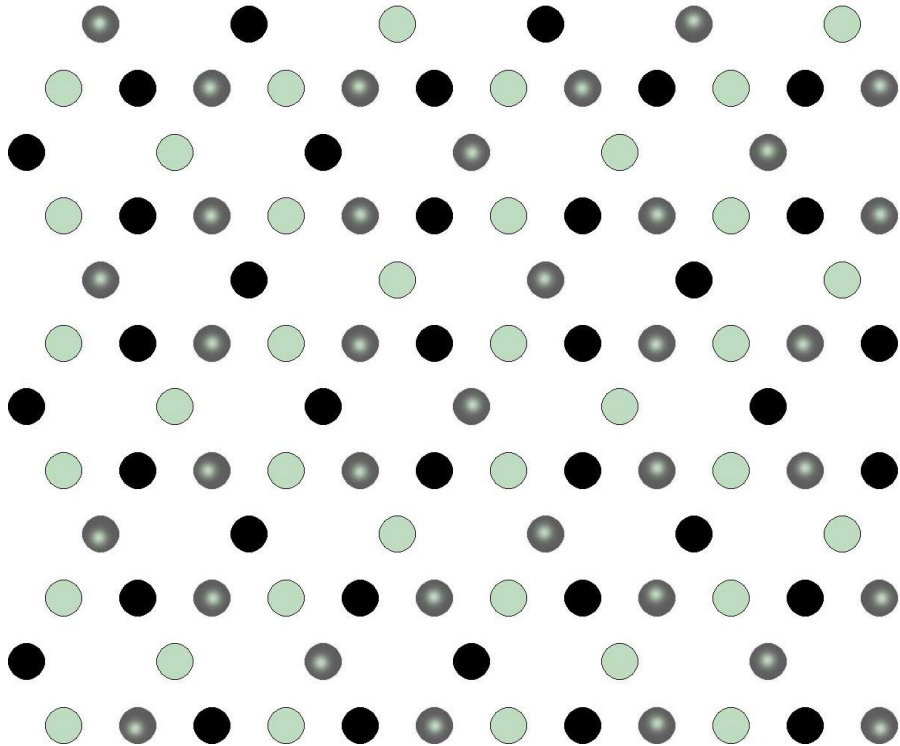

Figure 12. The magnetic structure of the Partially Disordered Phase.

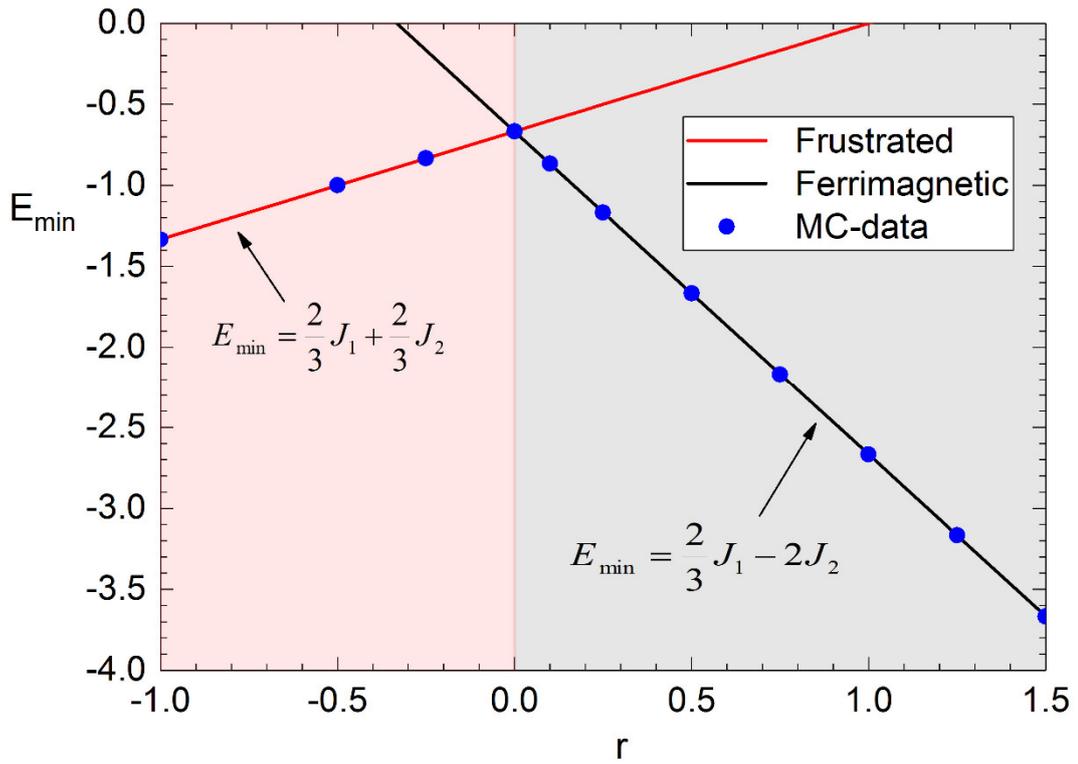

Figure 13. The ground state Energy for various r.

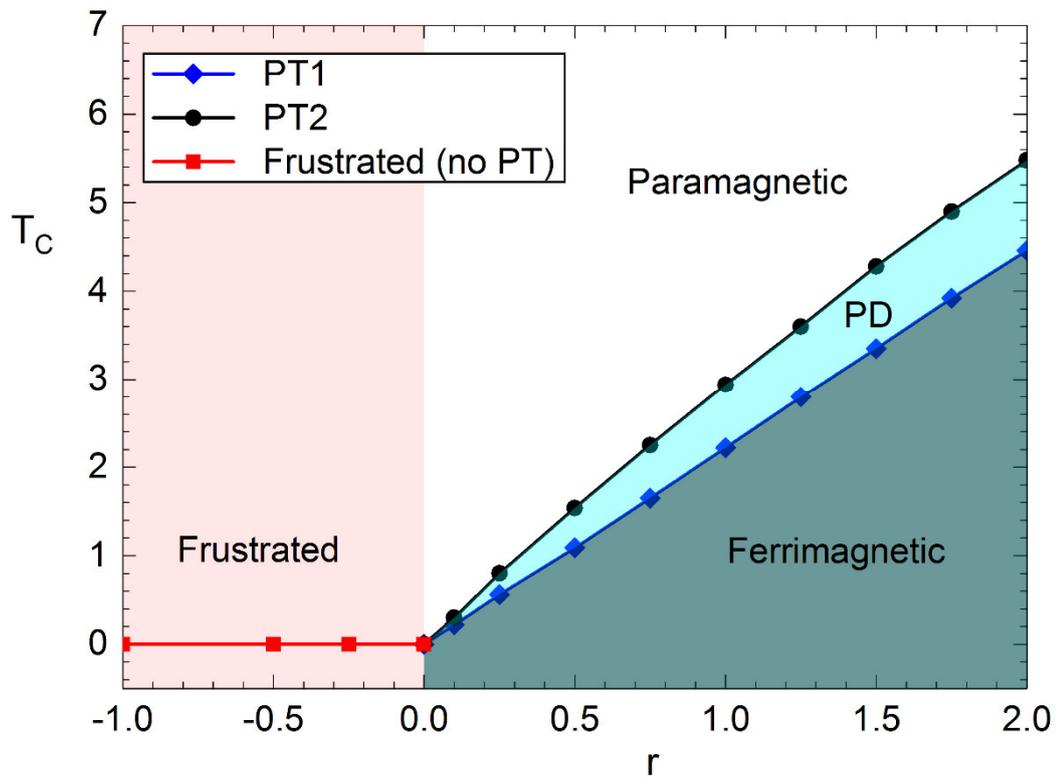

Figure 14. The Phase Diagram.